\documentclass[12pt]{article}
\usepackage{amsmath}
\usepackage{amscd}
\usepackage{amssymb}
\usepackage{array}

\begin{document}
\rightline{CERN-PH-TH/2004-033}

\rightline{IFIC/04-9}

\rightline{FTUV-04/0217}

\newcommand{\R}{\mathbb{R}}
\newcommand{\C}{\mathbb{C}}
\newcommand{\Z}{\mathbb{Z}}
\newcommand{\Hb}{\mathbb{H}}
\newcommand{\V}{\mathbb{V}}
\newcommand{\del}{\mathcal{D}}

\newcommand{\rSp}{\mathrm{Sp}}

\newcommand{\fsp}{\mathfrak{sp}}

\newcommand{\cL}{\mathcal{L}}

\newcommand{\id}{\relax{\rm 1\kern-.35em 1}}
\vskip 1cm

  \centerline{\LARGE \bf Axion gauge symmetries and  }
\bigskip
 \centerline{\LARGE \bf  generalized Chern--Simons terms    }
 \bigskip
 \centerline{\LARGE \bf in $N=1$  supersymmetric theories }

 \vskip 1.5cm
\centerline{L. Andrianopoli$^{\flat}$,  S.
Ferrara$^{\flat,\sharp}$ and M. A. Lled\'o$^{\natural}$.} \vskip
1.5cm

\centerline{\it $^\flat$ CERN, Theory Division, CH 1211 Geneva 23,
Switzerland} \centerline{{\footnotesize e-mail:
Laura.Andrianopoli@cern.ch, \; Sergio.Ferrara@cern.ch}}

\medskip

\centerline{\it \it $^\sharp$INFN, Laboratori Nazionali di
Frascati, Italy.}

\medskip
\centerline{\it $^\natural$
 Departamento de F\'{\i}sica Te\'orica,
Universidad de Valencia and IFIC}
 \centerline{\small\it C/Dr.
Moliner, 50, E-46100 Burjassot (Valencia), Spain.}
 \centerline{{\footnotesize e-mail: Maria.Lledo@ific.uv.es}}

\vskip 1cm

\begin{abstract}
We compute the extension of  the Lagrangian of $N=1$
supersymmetric theories to the case in which some axion symmetries
are gauged. It turns out that generalized Chern--Simons terms
appear  that were not considered in previous superspace
formulations of general $N=1$ theories.

Such gaugings appear in supergravities arising from flux
compactifications of superstrings, as well as from Scherk--Schwarz
generalized dimensional reduction in M-theory.

We also present the dual superspace formulation where axion chiral
multiplets are dualized into linear multiplets.

\end{abstract}

\vfill\eject

\section{Introduction}

In many models of superstring and M-theory compactified to $D=4$, it is
possible to obtain a scalar potential which stabilizes some of
the moduli (as well as matter) fields,  lifting therefore the
degeneracy of the moduli space of vacua. Examples of
this phenomenon are type II superstrings compactified on
orientifolds with NS and RR fluxes turned on \cite{frey} - \cite{dag}, or
Scherk-Schwarz generalized dimensional reductions \cite{ss} in M-theory.

In these theories the mass terms  arise through  a Higgs mechanism. The
supergravity description corresponds to the gauging of some axion
symmetries related to shifts of the scalar fields coming from wrapped RR forms or from the NS two-form
 B field in type II strings, and from the wrapped three-form in M-Theory.

In these gauge theories, generalized Chern--Simons terms emerge
\cite{dlv,tz,adfl,dst}.  If the gauge groups are abelian  they are
of the form
\begin{equation}
\frac 23c_{AB,\,C}\int A^A\wedge A^C\wedge dA^B,
\label{chernsimon}
\end{equation}
where
$c_{AB,\, C}$ are real constants with symmetries
\begin{equation}
c_{AB,\,C}=c_{BA,\,C}, \qquad
c_{AB,\,C}+c_{CA,\,B}+c_{BC,\,A}=0.
\label{symmc}
\end{equation}
 In the non abelian case an additional term is present
\begin{equation}
\frac 14c_{AB,\,C}f_{\ DE}^A\int A^D\wedge A^E\wedge A^C\wedge A^B,
\label{nonabelian}
\end{equation}
where $f_{\ DE}^A$ are the structure constants of the gauge group.

In theories arising from  type IIB compactification on $T_6/\Z_2$ orientifold \cite{fp,kst}, the
constants $c_{AB,\,C}$ are proportional to the RR and NS three-form fluxes $F^\alpha_{ABC}$ (with $A,B,C=1,\cdots ,6$;  $\alpha =1,2$),
and equation (\ref{chernsimon}) takes the form \cite{dfgtv,aft}
\begin{equation}
\frac 23F^\alpha_{ABC}\int A^A_\alpha\wedge A^B_\beta\wedge dA^C_\gamma \epsilon^{\beta\gamma},
\label{chernsimon2}
\end{equation}
Property (\ref{symmc}) is understood in these theories from the
fact that, in (\ref{chernsimon2}),
$$
F_{ABC}^{[\alpha}\epsilon^{\beta\gamma]}=0
$$
where the bracket $[\cdots]$ stands for complete antisymmetrization
of the indices.

 If we perform a  Scherk--Schwarz dimensional reduction to
M-theory with  Scherk--Schwarz phase matrix $M^A_{\ B}$, the
constants $c_{AB,C}$ come from the 5-d Chern--Simons form
\cite{sv,adfl}
$$d_{ABC}A^A \wedge F^B\wedge F^C\qquad (A,B,\dots =1,\cdots ,27)$$
and are given by
$$c_{AB,C}=d_{ABD}M^D_{\ C}.$$
In this case the condition (\ref{symmc}) is a consequence of the fact that,
in $N=8$ $d=5$ supergravity, $d_{ABC}$ is an $E_6$ invariant tensor.

 Another
instance where a particular form of such terms appears is in
deconstructed supersymmetric $U(1)$ gauge theories \cite{dfp},
where it arises for cancellation of mixed $U(1)$ anomalies
\cite{dfp,akt}.

The occurrence of such terms was studied for $N=2$ in Ref.
\cite{dlv}, but for a matter coupled $N=1$ supersymmetric gauge
theory they have not been considered previously.

It is the aim of the present investigation to give such completion
for the $N=1$ case. We will consider only the abelian gauge
groups, as for example the groups of axion shift symmetries.

Let $V^A$ be the superfield vector potentials and $W^A_\alpha=\bar
\del^2\del_\alpha V^A$ denote the chiral supersymmetric field
strengths, with  $A=1, \dots n_v$, where $n_v$ denotes  the number
of vectors undergoing the gauging and $\alpha$ is a spinor index.
The gauge transformations depend on chiral superfield parameters
$\Lambda^A$,

$$V^A\longrightarrow V^A+\Lambda^A+\bar\Lambda^A, \qquad
W^A\longrightarrow W^A.$$

 Let the kinetic term of the vectors in the
Lagrangian be written as \cite{cfgv}
\begin{equation}
\int d^2\theta f_{AB}W^AW^B \;+\; \mbox{h. c.},
\label{vectlag}
\end{equation}
(we have suppressed the contracted spinor indices) where the
matrix $f_{AB}$ is a holomorphic function of the scalar fields. We
assume that under a gauge transformation (we will justify later
this assumption)  the kinetic matrix of the vectors transforms as
$$\delta_\Lambda f_{AB}=c_{AB,C}\Lambda^C$$ with $c_{AB,C}$ the real constants appearing in
 (\ref{chernsimon})). Gauge invariance is achieved
because of the presence of the generalized Chern--Simons terms
which involve only the vector fields.

The axionic chiral multiplets can be dualized into linear
multiplets \cite{cfv}. The dual lagrangian exhibits Green-Schwarz
couplings of the linear multiplets. Let $b_i$ denote the two-form
fields dual to the axion fields $S^i$ and  $F^A$  the field
strengths of the gauge potentials. If we write the gauge
transformations of the dual axion multiplets as
$$\delta S^i=M_A^i\Lambda^A,$$
then the G-S coupling terms are the supersymmetric extensions of the
bosonic dual terms
 $$
M_A^ib_i\wedge F^A.
$$

\bigskip

The paper is organized as follows. In section 2 we recall the
symplectic action of the $\sigma$-model isometries on vector
fields as duality rotations and the need to introduce generalized
Chern--Simons terms. In section 3 we derive the dual lagrangian
with linear multiplets, whose physical bosonic components are
antisymmetric tensors. Conclusions and outlooks are given in
section 4. An appendix with some useful formulae is included.

\section{Dualities, axionic symmetries and Chern--Simons terms\label{dualities}}

The standard form of the Lagrangian density in $N=1$
supersymmetric gauge theories is \cite{cfgv}
\begin{equation}
\int d^4\theta
K(S, \bar Se^V)+\left[\int d^2\theta \left(f_{AB}(S)W^AW^B + P(S)\right) \,+\, \mbox{h.
c.}\right],
\label{standardlagrangian}
\end{equation}
where $K$ and $P$ (the K\"ahler potential, and the superpotential
respectively) are gauge invariant. The matrix $f_{AB}$ is a chiral
superfield, symmetric in the indices $A,B$, transforming in the
 twofold symmetric tensor product of
the coadjoint representation of the gauge group, to make the action
 gauge invariant.

From the structure of the vector couplings \cite{gz} it follows
that $f_{AB}$ may have a more general transformation rule. The
invariance of the system of field equations plus Bianchi
identities \cite{cdfv} allows a transformation of $f_{AB}$ in
terms of a matrix of $\rSp(2n_v,\R)$
\begin{equation}
\begin{pmatrix}A& B\\C& D\end{pmatrix}
 \qquad A^tC, B^tD\;\mbox{symmetric},\quad
A^tD-C^tB=1
\label{sympl}
\end{equation}
of the form
\begin{equation}
if'=(C+Dif)(A+Bif)^{-1}.
\label{convention}
\end{equation}
The
transformations of $\rSp(2n_v,\R)$ mix electric and magnetic field
strengths. When a subgroup of this symmetries becomes a local
(gauge) symmetry, it must act on the gauge potentials so it must
be {\it electric}; this means that necessarily $B=0$. This implies
$A^t=D^{-1}$ and
\begin{equation}
if'=CA^{-1}+{(A^t)}^{-1}ifA^{-1},
\label{transff}
\end{equation}
while the gauge vectors transform simply as $V'=AV$. Nevertheless,
the gauge group may have, as embedded in $\rSp(2n_v,\R)$, $C\neq
0$. The gauge group corresponds, in the scalar manifold, to a subset
of the isometry group that has been gauged. A gauge transformation
will result  in a transformation of $f$ (as a function of the
scalar fields) of the type (\ref{transff}). If the symplectic
transformation is constant, this will result in a change of the
Lagrangian as a total derivative\footnote{Note that the convention
taken in (\ref{convention}) is essential for this to be true}, but
if the transformation depends on local parameters, then the
Lagrangian (\ref{standardlagrangian}) must be modified to achieve
gauge invariance.

One can gauge  abelian groups, which have $B=0$, $A=1$ and $C\neq
0$. They have a non trivial action on the scalar fields. Non
abelian gaugings ($A\neq 0$) with $C\neq 0$ are also possible but will not be
considered here.

\bigskip

 From now on we will assume that the gauge group is abelian.
 We  assume that we can choose local coordinates in the scalar manifold
 in such way that  the set of chiral multiplets ($n_c$)   can be
 split
  in two sets,

   $$\{S^i\}_{i=1}^n \qquad  \mbox{and} \qquad \{T^a\}_{a=1}^m,  \qquad n+m=n_c,$$
  in such way that  only the multiplets $S^i$ (together with the vector potentials) transform
 under the gauge group. We have
\begin{equation}
\delta_\Lambda T^a=0, \qquad \delta_\Lambda S^i=M^i_A\Lambda^A,
 \qquad \delta_\Lambda V^A=\Lambda^A+\bar \Lambda^A.
\label{gaugetrasf1}
\end{equation}
 Also, we assume that the matrix $f_{AB}$ is of the form
 $$f_{AB}(S,T)= d_{ABi}S^i +\tilde f_{AB}(T).$$
 Then
\begin{equation}
\delta_\Lambda f_{AB}(S,T)=c_{AB,\,C}\Lambda^C,
\label{gaugetrasf2}
\end{equation}
with
\begin{equation}
c_{AB,\,C}=d_{ABi}M^i_C.
\label{c=dm}
\end{equation}
 We will see later that the
lagrangian can be made gauge invariant if   $d_{ABi}M^i_C$ are such that
the second property in (\ref{symmc}) holds.

The combination $S^i+\bar S^i-M_A^iV^A$ is gauge invariant, and
the kinetic term for the chiral fields is of the form
$$\int d^4\theta K(T, S^i+\bar S^i- M_A^iV^A).$$ (The fields $S^i$
in this expression are the logarithms of what in
(\ref{standardlagrangian}) was denoted by $S$).

The relevant vector kinetic term is
\begin{equation}
\int d^2\theta d_{ABi}S^iW^AW^B \;+\;\mbox{h. c.},
\label{vectlagr}
\end{equation}
and its gauge variation is
\begin{equation}
\int d^2\theta c_{AB,\,C}\Lambda^CW^AW^B \;+\;\mbox{h.
c.}.
\label{variation}
\end{equation}
This term is  a total
derivative if $\Lambda^C(x,\theta)$ is an imaginary constant. Otherwise,
to cancel this variation we must add a generalized Chern--Simons
term, which can be constructed with the Chern--Simons multiplet
introduced in Ref. \cite{cfv}. This term is
\begin{equation}
-\frac 2 3c_{AB,\,C}\int d^4\theta V^C\Omega^{AB}(V),
\label{extraterm}
\end{equation}
 with
\begin{equation}\Omega^{AB}(V)= \del^\alpha V^{(A}W_\alpha^{B)}+ \bar \del_{\dot\alpha} V^{(A}W^{\dot\alpha B)}+
 V^{(A}\del^\alpha W_\alpha^{B)},
\end{equation}
 (the parenthesis $(\cdots)$ stands for symmetrization in the indices). Note that $\Omega$ is real; in particular, the last term is
 real because of the Bianchi identity
 $$\del^\alpha W_\alpha=\del_{\dot\alpha}W^{\dot\alpha}=\overline{\del^\alpha W_\alpha}.$$
 Under a gauge transformation the Chern--Simons multiplet
 transforms as
 $$\delta_\Lambda\Omega^{AB}(V)=\del^\alpha(\Lambda^{(A}W_\alpha^{B)})+\bar
  \del_{\dot \alpha}(\bar\Lambda^{(A}\bar W^{\dot\alpha B)}).$$
 Using the property (proven in the Appendix)
 $$V^{(C}\Omega^{AB)}=\frac 16\del^\alpha\left(V^AV^BW^C_\alpha+V^AV^CW^B_\alpha+V^BV^CW^A_\alpha\right) + \mbox{h. c.},$$
 the part  in  (\ref{extraterm}) which is symmetric in $(A,B,C)$ gives zero
 contribution to the action, being  a total space-time derivative.
 This means that if $c_{(AB,\,C)}\neq 0$ the variation
 (\ref{variation}) cannot be completely cancelled by a term like
 (\ref{extraterm}). So we require that $c_{(AB,\,C)}= 0$ as a
 consistency condition for gauge invariance.

 The gauge variation of (\ref{extraterm}) is (see Appendix)
 $$-c_{AB,\,C}\left(\int d^2\theta\Lambda^CW^AW^B \;+\;\mbox{h.c.}\right),$$
 and we see that it cancels exactly the gauge variation of the
 vector kinetic term (\ref{variation}),
so that the Chern--Simons-completed vector lagrangian
$$ -\frac 23 c_{AB,C}\int d^4\theta \Omega^{AB}V^C + \left(\int d^2\theta f_{AB}W^AW^B \;+\; \mbox{h. c.}\right)$$
is gauge-invariant.

 This is in agreement with what was found in
 Ref. \cite{dlv} for the $N=2$ case.

\bigskip

Let us further observe that, in the Wess--Zumino gauge,
 the component expression of the Chern--Simons action (\ref{extraterm}) contains, beyond the bosonic  contribution (\ref{chernsimon}),
the extra term
$$
c_{AB,C}\bar \lambda^A \gamma^\mu \gamma_5 \lambda^B A^C_\mu .
$$
This is needed in order to make gauge-invariant the fermionic
contribution in (\ref{vectlagr})  containing ${\rm Im} \phi^i$
($\phi^i = S^i |_{\theta=0}$)
$$
 d_{AB,i}\, {\rm Im} \phi^i \partial_\mu \left(\bar \lambda^A \gamma^\mu \gamma_5 \lambda^B\right)
$$
which then becomes, using (\ref{c=dm}),
$$
 - \, d_{AB,i}\left( \partial_\mu {\rm Im} \phi^i - M^i_{\ C} A_\mu^C\right)\bar \lambda^A \gamma^\mu \gamma_5 \lambda^B  .
$$

\section{Dual form of the lagrangian}
The lagrangian studied in the previous section can be dualized by
replacing the chiral multiplets $S^i$ by the dual linear
multiplets $L_i$. These are real multiplets satisfying the
constraint $\del^2L_i=\bar \del^2L_i=0$. In order to perform the
dualization we  introduce the real superfield $U^i$. The lagrangian
connecting the two theories is

\begin{eqnarray}
\cL&=&\int d^4\theta \left[K(T, U^i-M^i_AV^A) -L_iU^i + \left( d_{ABi}U^i
 -\frac 23 c_{AB,\,C} V^C\right)\Omega^{AB}\right]
 +\nonumber\\
&+&\left[\int d^2\theta \tilde
f_{AB}(T)W^AW^B\;+\;\mbox{h. c.}\right],\label{mothertheory}
\end{eqnarray}
with
$c_{AB,\,C}=d_{ABi}M^i_C$. The original Lagrangian is obtained by
varying (\ref{mothertheory}) with respect to $L_i$, which gives
$U^i=S^i+\bar S^i$  (notice that $L_i$ is not unconstrained), and
substituting back in $\cL$.

The dual Lagrangian instead is obtained by varying with respect to
$U^i$ and substituting the equation obtained in $\cL$. Let us
define $\tilde U^i=U^i-M^i_AV^A$. Then, the relevant terms in
(\ref{mothertheory}) become
$$
\int d^4\theta \left[K(T, \tilde
U^i)-\tilde U^i(L_i-d_{ABi}\Omega^{AB})-L_iM^i_AV^A+\frac 13
c_{AB,\,C}V^C\Omega^{AB}\right].
$$
Solving
$$
\Psi_i\equiv\frac{\partial K(T, \tilde U^i)}{\partial\tilde
U^i}-L_i+d_{ABi}\Omega^{AB}=0
$$
one gets
$$\int d^4\theta \left[\Phi(T, L_i-d_{ABi}\Omega^{AB})-
L_iM^i_AV^A+\frac 13 c_{AB,\,C}V^C\Omega^{AB}\right],
$$
where
$$
K(T, \tilde U^i)-\tilde U^i(L_i-d_{ABi}\Omega^{AB})=\Phi(T,
L_i-d_{ABi}\Omega^{AB})\quad \mbox{at} \quad \Psi_i=0.
$$

The gauge transformation of $L_i$ are
$$
\delta L_i=d_{ABi}\left[\del^\alpha\left(\Lambda^{A}W^{B}_\alpha\right)+
\bar\del_{\dot\alpha}\left(\bar\Lambda^{A}W^{B\dot\alpha}\right)\right].
$$
Notice
that the variation of the Green--Schwarz term is now cancelled by
the variation of the generalized Chern--Simons term (which has a different coefficient
with respect to the dual formulation).

\section{Conclusions}
In this investigation we have given the superfield expression of the
$N=1$ lagrangian with gauged axion symmetries.

The lagrangian requires new coupling terms which were not present
in the standard formulation because there it was assumed that the
gauge transformations changed the  kinetic matrix of the vectors
as
$$
f'= (A^t)^{-1}f A^{-1},
$$
where $A$ is the adjoint action of the
fields. It is interesting to observe that not all the axion gauge
symmetries can be gauged, but only those for which the expression
$$
d_{ABi}Mî_C=c_{AB, \, C}\qquad \mbox{satisfies}\qquad c_{(AB, \,
C)}=0.
$$
In fact, the variation of the action with the term (\ref{extraterm}) included is
$$
c_{(AB, \,
C)}\int d^2\theta \Lambda^CW^AW^B\;+\;\mbox{h. c.}
$$
and it vanishes only when the above consistency condition is fulfilled.
The simplest case where $c_{(AB,C)} \neq 0$ is when we have only
one axion $S$ with coupling
$$SW^AW^B\delta_{AB}.$$
Under the axion symmetry $S\rightarrow S+\Lambda$ the lagrangian is not gauge-invariant at the classical level, rather it can be used to cancel (one-loop) quantum anomalies \cite{dsw,dfkz,ds}.

It is possible to extend the present analysis to the
supergravity case and  to non-abelian axion symmetries. Such
cases are incountered in Scherk--Schwarz M-theory compactifications
and in type IIB supergravity compactifications in the presence of
fluxes.

In the non-abelian case ($A\neq \id$ in equation (\ref{sympl})), the coefficients $c_{AB,C}$ must satisfy the extra condition \cite{dlv}
\begin{equation}
f^D_{\ E(B} c_{A)D,F} - f^D_{\ F(B} c_{A)D,E} + \frac 12 f^D_{\ EF} c_{AB,D}=0.
\label{extracondition}
\end{equation}
This follows from the fact that for a vector transforming as
$$
\delta {\mathrm{V}}_a = t_{a \ \ A}^{\ b}\Lambda^A {\mathrm{V}}_b  + C_{aA}\Lambda^A
$$
the closure of the gauge algebra requires a cocycle condition on the coefficient $C_{aA}$ \footnote{We thank R. Stora for enlightening discussions on this issue.}
$$
t_{a \ \ A}^{\ b}C_{bB} - t_{a \ \ B}^{\ b}C_{bA} - f_{AB}^{\ \ C} C_{aC}=0.
$$
The condition (\ref{extracondition}) on the coefficients $c_{AB,C}$ is just the above relation,
when specified to the twofold symmetric product of the coadjoint representation.

In the Wess--Zumino gauge, the supersymmetric version of the non abelian completion (\ref{nonabelian}) is
$$
c_{AB,C} f^B_{\ PQ} V^C \del^\alpha V^A \bar \del^2 \left( \del_\alpha V^P V^Q\right) + \mbox{h. c. }
$$

\section*{Appendix: Some useful relations}

Consider the superfield:
\begin{equation}
\Omega^{AB}={\mathcal{D}}^\alpha V^{(A}W_\alpha^{B)} + \bar{\mathcal{D}}_{\dot\alpha} V^{(A}\bar W^{\dot\alpha B)} + V^{(A}{\mathcal{D}}^\alpha W_\alpha^{B)}
\end{equation}
where $(\cdots)$ stands form complete symmetrization in the
indices.

$\Omega^{AB}$ is real thanks to the property
\begin{equation}
\del^\alpha W_\alpha = \bar\del_{\dot\alpha}\bar W^{\dot\alpha}
\label{real}
\end{equation}
and it satisfies:
\begin{equation}
\bar\del^2 \Omega^{AB} =  W^{(A\alpha} W_{\alpha}^{ B)}
\label{d2o}
\end{equation}

Consider now the superfield $\Omega^{AB}V^C$.
Its totally symmetric part can be written as a total derivative
\begin{eqnarray}
\Omega^{(AB}V^{C)}
& = & \frac 12\del^\alpha\left( V^{(A}V^BW_\alpha^{C)}\right) + \mbox{h. c. }\nonumber\\
&=& \frac16 \del^\alpha \left(V^AV^BW_\alpha^C + V^B V^CW_\alpha^A + V^AV^CW_\alpha^B\right) + \mbox{h. c. }
\end{eqnarray}
so that such a lagrangian term does not contribute to the action:
\begin{equation}
\int d^4 x \int d^4\theta \Omega^{(AB}V^{C)} =0.
\end{equation}
Indeed, using (\ref{real}), we have:
\begin{eqnarray}
\Omega^{(AB}V^{C)}&=& \del^\alpha V^{(A} W_\alpha^B V^{C)} +\frac 12  V^{(A}\del^\alpha W_\alpha^B V^{C)} + \mbox{h. c. } \nonumber\\
& =& \frac 12\del^\alpha\left( V^{(A}V^B\right)W_\alpha^{C)}  +\frac 12  V^{(A}\del^\alpha W_\alpha^B V^{C)} + \mbox{h. c. }\nonumber\\
&= & \frac 12\del^\alpha\left( V^{(A}V^BW_\alpha^{C)}\right) + \mbox{h. c. }
\end{eqnarray}
\bigskip

{\bf The gauge invariance}
\bigskip

\noindent
We explicitely show here, for the abelian case,
 that the lagrangian (\ref{standardlagrangian}) completed with the Chern--Simons term (\ref{chernsimon}) is gauge-invariant.

Indeed, from (\ref{gaugetrasf1}) we have
$$\delta_\Lambda \Omega^{AB}= \del^\alpha\left(\Lambda^{(A}W^{B)}_\alpha\right) + \mbox{h. c. }. $$
Then
\begin{eqnarray}
&\delta_\Lambda \left(c_{AB,C}\int d^4\theta
\Omega^{AB}V^C\right)= c_{AB,C}\int d^4\theta
\left[\Omega^{AB}\Lambda^C+\del^\alpha
\left(\Lambda^A W^B_\alpha\right)V^C\right] + \mbox{ h. c.}=\nonumber\\
&=c_{AB,C}\int d^2\theta  \left(W^AW^B\Lambda^C-\Lambda^A W^B
W^C\right)+ \mbox{ h. c.} \label{csvariation}
\end{eqnarray}
where we have used (\ref{d2o}) and the notation
$$W^AW^B = W^BW^A \equiv W^{\alpha A} W^B_\alpha = - W^A_\alpha W^{\alpha B}.$$
However, due to equations (\ref{symmc}), we have
\begin{eqnarray*}
c_{AB,C}\Lambda^A W^BW^C &=& (-c_{AC,B}- c_{BC,A})\Lambda^A W^BW^C \\
&=& -c_{AB,C}\Lambda^A W^BW^C - c_{AB,C}\Lambda^C W^AW^B
\end{eqnarray*}
so that
\begin{equation}
c_{AB,C}\Lambda^A W^BW^C = -\frac 12  c_{AB,C}\Lambda^C W^AW^B \label{simpl}
\end{equation}
Using (\ref{simpl}) in (\ref{csvariation}) we get
\begin{equation}
\delta_\Lambda \left(c_{AB,C}\int d^4\theta \Omega^{AB}V^C\right) =\frac 32 c_{AB,C}\int d^2\theta\Lambda^C W^AW^B\;+\; \mbox{h. c.}
\label{csvar}
\end{equation}
On the other hand, from (\ref{gaugetrasf2}), the gauge transformation of the vector kinetic term (\ref{vectlag}) is
\begin{equation}
\delta_\Lambda \left(\int d^2\theta f_{AB}W^AW^B \;+\; \mbox{h. c.}\right) =
 c_{AB,C}\int d^2\theta \Lambda^C W^AW^B \;+\; \mbox{h. c.}
\label{vectlagrasf}
\end{equation}
so that the Chern--Simons-completed vector lagrangian
$$ -\frac 23 c_{AB,C}\int d^4\theta \Omega^{AB}V^C + \left(\int d^2\theta f_{AB}W^AW^B \;+\; \mbox{h. c.}\right)$$
is gauge-invariant.

\section*{Acknowledgements}
S.F. acknowledges interesting discussions with E. Dudas, E.
Kiritsis and C. Kounnas. L. A. wants to thank the Departamento de
F\'{\i}sica Te\'orica, Universidad de Valencia for its kind
hospitality during the realization of this work. The work of S. F
has been supported in part by the D.O.E. grant DE-FG03-91ER40662,
Task C.

\end{document}